\documentclass[]{aastex631}
\usepackage{appendix}
\usepackage{datatool,booktabs}
\usepackage{subfigure} % 

\usepackage{amsmath}

\def\aap{Astron.\ Astrophys.\ }

\def\apjl{Astrophys.\ J.\ Lett.\ }
\def\apjs{Astrophys.\ J.\ Supp.\ }

\begin{document}

\title{ Solar modulation of AMS-02 daily proton and Helium fluxes with modified  force field approximation models }

\author{Cheng-Rui Zhu}
\email{zhucr@ahnu.edu.cn}  
\affiliation{Department of Physics, Anhui Normal University, Wuhu 241000, Anhui, China }
\affiliation{Key Laboratory of Dark Matter and Space Astronomy, Purple
Mountain Observatory, Chinese Academy of Sciences, Nanjing 210008, Jiangsu, China }

\author{Mei-Juan Wang}
\affiliation{Department of Physics, Anhui Normal University, Wuhu 241000, Anhui, China }

\begin{abstract}

 As galactic cosmic rays propagate through the turbulent plasma environment within the heliosphere, they undergo a process of diffusion, drift and energy loss, leading to a notable reduction in their flux. This is the solar modulation impact. Recently, the cosmic-ray experiment AMS-02 published  daily fluxes of proton and Helium for the period from May 20, 2011 to October 29, 2019 in the rigidity interval from about 1 to 100 GV, exhibiting fine time structures that correlate with solar wind properties on daily basis. In this work, we employ three different modified force field approximation models to fit the data.  By fitting to the daily proton and Helium fluxes, we get the time series of solar modulation potential. We find good agreement of data and model predictions for both proton and Helium with the same parameters in two modified force field approximation models. The results in this study verify that the modified FFA model is a valid parametrization of the GCR spectrum also at daily time scales.
 %The results verify the modified force field approximation of Parker’s solar modulation equation on daily basis. 

\end{abstract}

%%%%%%%%%%%%%%%%%%%%%%%%%%%%%%%%%%%%%%%%%%%%%%%%%%%%%%%%%%%%%%

\section{Introduction}

The galactic cosmic rays (GCRs) are charged, energetic particles coming from  cosmic accelerators such as supernova remnants, and then propagate diffusively in the Galactic random magnetic field \citep{Moskalenko_1998}. When entering the heliosphere, they encounter a turbulent solar wind with an embedded heliospheric magnetic field (HMF), leading to the local interstellar spectrum (LIS) of GCRs changing into the top-of-atmosphere spectrum (TOA). \citep{1971RvGSP...9...27J,Potgieter2013}. This is the solar modulation process of GCRs. Since most of our experiments to detect GCRs are conducted within the heliosphere, this process limits our understanding of GCRs. Consequently, the solar modulation effects  are crucial to  our understanding of the injection and propagation parameters of cosmic rays, dark matter indirect measurement and the diffusion theory in the galaxy and heliosphere \citep{2003AdSpR..32..549B,2007ARNPS..57..285S,2013A&ARv..21...70B,2012CRPhy..13..740L,Yuan:2014pka,2017ApJ...840..115B,2017PhRvL.118s1101C,Tomassetti:2017hbe,Yuan_2018,2022PhRvL.129w1101Z}.

The basic transport equation (TPE) was derived by \cite{1965P&SS...13....9P,1967ApJ...149L.115G} to  describe the GCRs’ transport processes in the heliosphere. \citep{1968ApJ...154.1011G} also derived an approximate solution to this TPE, the so called force-field approximation (FFA), which had been widely used  as it is simple and enough to explain most of the observations. However, with the advancement of observational technology, the force field approximation increasingly reveals its insufficiencies. 
%The numerical models  or modified FFA models are used to solve this problem.

With the successful progress of experiments such as PAMELA, AMS-02,  DAMPE \citep{2011Sci...332...69A,2017PhRvL.119y1101A,cite-key}, study on cosmic rays has entered an era of high precision. Especially, AMS-02 not only provides high-precision energy spectra for various cosmic ray particles but also offers the temporal evolution of multiple cosmic ray spectra, including protons, Helium, deuterium, He$^3$, He$^4$, electrons, and positrons. These data have greatly enhanced our understanding of solar modulation. As the FFA cannot explain the data features, people have adopted various methods, including numerical  models  and modified FFA models to solve this problem. \cite{PhysRevLett.121.251104,Luo_2019,Corti_2019,Song_2021,2022PhRvL.129w1101Z,2022PhRvD.106f3006W} reproduced the AMS observations using a one-dimensional or a three-dimensional numerical model  to solve the Parker equation.
{Several methods have been proposed to expand the  FFA \citep{2016ApJ...829....8C,2016PhRvD..93d3016C,2017PhRvD..95h3007Y,2017JGRA..12210964G,Zhu:2020koq,2021ApJ...921..109S,Li_2022,Cholis_2022,PhysRevD.109.083009,zhu2024}  to account for the differences in observed and predicted GCR spectra.}

 In the past, we can only study the solar modulation on daily basis with neutron monitors (NMs) \citep{1999CzJPh..49.1743U,Ghelfi_2017}. Recently,  the AMS-02 published the daily fluxes of proton and Helium \citep{AMS:2021qln,AMS:2022ojy} from 1 GV to 100 GV providing new  opportunities to study the solar modulation on daily basis based on space-based detector. In this work, we will study these daily fluxes using three modified FFA models that all exhibit rigidity-dependent solar modulation potential.

\section{Methodology}
\subsection{Modified force-field approximation models}

%A full solution to the problem related to CRs transport in the heliosphere is a complicated task and requires sophisticated 3D time-dependent self-consistent modelling. However, the problem can be essentially simplified by the force-field approximation (FFA).
%The force-field approximation (FFA), derived from the Parker transport equation (TPE) \citep{1965P&SS...13....9P} with several approximations, is widely used due to its simplicity. Usually, the FFA requires the  quasi-steady changes and azimuthal symmetry. The azimuthal symmetry requires times longer than the solar-rotation period of $\approx 27$ days synodic \citep{Usoskin_2023}. However, when cosmic rays arrive the Earth, it has already been about a year since they enter the heliosphere. Therefore, in most cases, they meet the assumption of azimuthal symmetry. When cosmic rays propagate around the Earth, they are not influenced by other areas of the heliosphere, it still meet the azimuthal symmetry assumption. So the azimuthal symmetry assumption is also Validated for the daily CRs fluxes.
A full solution to the problem related to CRs transport in the heliosphere is a complicated task and requires sophisticated 3D time-dependent self-consistent modelling \citep{Usoskin_2023}. However, the problem can be essentially simplified by the force-field approximation (FFA). The FFA, derived from the Parker transport equation (TPE) \citep{1965P&SS...13....9P} with several approximations, is widely used due to its simplicity. 
Usually, the FFA requires the  quasi-steady changes,  spherical symmetry, etc, witch are apparently invalid for short time scales. In fact, they are not fully valid even for the regular condition \citep{2003JA010098}. However, the force-field formalism was found to provide a very useful and comfortable mathematical parametrization of the GCR spectrum even during  a major Forbush decrease (FD), irrespective of the (in) validity of physical assumptions behind the force-field model. So one can still benefit from the simple parametrization  offered by the FFA for practical uses other than studying the physics of solar modulation, even if the FFA is not a physically motivated solution to the solar modulation problem at daily time scales. \citep{USOSKIN20152940}.
In the FFA model, the TOA flux is related with the LIS flux as
\begin{equation}\label{force_filed}
J^{\rm TOA}(E)=J^{\rm LIS}(E+\Phi)\times\frac{E(E+2m_p)}
{(E+\Phi)(E+\Phi+2m_p)}, 
\end{equation}
where $E$ is the kinetic energy per nucleon, $\Phi=\phi\cdot Z/A$ with $\phi$ being the solar modulation potential, $m_p=0.938$ GeV is the 
proton mass, and $J$ is the differential flux of GCRs. The only
parameter in the force-field approximation is the modulation potential $\phi$.

With the development of instruments, the observational data shows that we cannot fit the cosmic ray flux well with just one parameter, therefore, an rigidity-dependent solar modulation potential is needed \citep{2017JGRA..12210964G,SIRUK20241978}. Several methods have been proposed to describe the rigidity-dependent of solar modulation potential, such as \cite{2021ApJ...921..109S,Cholis_2022,PhysRevD.109.083009,zhu2024}. In this work we describe the rigidity-dependent of solar modulation potential with three different modified FFA models. First, the solar modulation potential in Zhu's model \citep{zhu2024} is shown as
\begin{equation}\label{sigmoid}
\phi(R)_{Zhu}  = \phi_l +\left (\frac{\phi_h-\phi_l}{1+e^{(-R+R_b)}} \right )
\end{equation}
where $\phi_l$ is the solar modulation potential for the low energy, and $\phi_h$ is for the high energy, $e$ is the natural constant, $R$ is the rigidity and $R_b$ is the break rigidity. This model is developed from the model in \cite{2016ApJ...829....8C}.
%, and we do not used the normalization index \citep{Zhu:2020koq,PhysRevD.109.083009} here which  may be related to the drift effect.

Second,  Cholis’ model from \cite{2016PhRvD..93d3016C,PhysRevD.106.063021} is described as
\begin{equation}\label{Cholis}
 \phi (R)_{Cholis} = \phi_0 + \phi_1 \left(  \frac{1+(R/R_0)^2}{\beta(R/R_0)^3}   \right) .
\end{equation}
Here, $\beta$ is the  ratio between the particle speed and the speed of light. $\phi_0$, $\phi_1$ and $R_0$ are the free parameters to be fitted. In the original Cholis model, $\phi_0$ and $\phi_1$ are parameters that depend on the magnetic field intensity (B) and tilt angle ($\alpha$). We allow these parameters to vary freely in order to obtain better fitting results.

Third,  Long's model from \cite{Kuhlen_2019,PhysRevD.109.083009} is described as 
\begin{equation}\label{Long}
 \phi (R)_{Long} = \phi_0 + \phi_1 ln(R/R_0), 
\end{equation}
with 
\begin{equation}\label{force_filed3}
\begin{aligned}
J^{\rm TOA}(E)=&J^{\rm LIS}(E+ \Phi(R))
\times \frac{E(E+2m_p)}{(E+\Phi(R))(E+\Phi(R)+2m_p)}exp(-g\frac{10R^2}{1+10R^2}\phi(R))).
\end{aligned}
\end{equation}
Here, $\phi_0$, $\phi_1$ and $g$ are the free parameters to be fitted with $\Phi(R) = \phi(R) \cdot Z/A $.  
%\cite{2021ApJ...921..109S} presents Shen's model. However, to achieve better fitting results for different cosmic-ray species, it is necessary to modify the model to be rigidity dependent rather than energy dependent. Therefore, we do not employ their model in this work.

\subsection{LIS of proton and Helium}

\begin{figure}[!ht]
    \centering
    \includegraphics[scale=0.6]{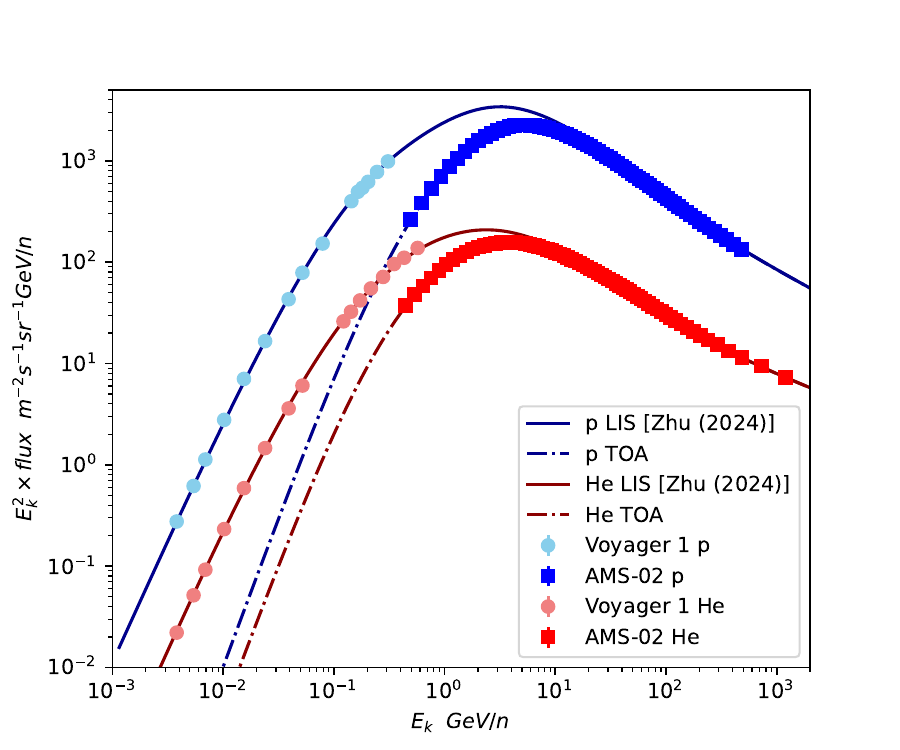}
    \caption{ LIS fluxes (lines) and TOA fluxes (dashed lines
   ) of p and He from \cite{zhu2024}, multiplied by 
    $E_k^{2}$, compared with the measurements (colorful points) 
    of Voyager 1 \citep{2016ApJ...831...18C}, and AMS-02 \citep{AGUILAR20211}.
  }
    \label{fig:LIS}
\end{figure}

There are different p and He LIS assumption in the literature, such as \cite{Potgieter2014,2016A&A...591A..94G,2016ApJ...829....8C,BOSCHINI20182859,zhu2024}. In this work, we take the LIS from our previous work \cite{zhu2024}, where we get the LIS with the  cubic spline interpolation method, with the highest-order of polynomial of three, and we still assume Z/A=1/1 for proton and Z/A = 2/4 for He. The LISs for proton and Helium are shown in Fig. \ref{fig:LIS}. 
%Different LIS will cause different $\phi$ series, however the modulation potential can be converted from one LIS to the other using a linear relation \citep{Asvestari2017}.

%In order to study the time-dependent solar modulation effects, we need the LIS of proton and Helium. Usually power-law or broken power-law functions are employed to fit the GCRs data \citep{2006AdSpR..37.1727O,2014A&A...566A.142Y,2016ApJ...829....8C}. 
%If the observational data cover a wide enough energy range, one can instead use a non-parametric method by means of spline interpolation of GCR fluxes among a few knots \citep{2016A&A...591A..94G,Zhu:2018jbk}, which can avoid the bias from the cosmic ray injection and propagation model. The spline interpolation is a way to obtain an approximate function smoothly passing through a series of points using piecewise polynomial functions. We use the cubic spline interpolation here, with the highest-order of polynomial of three. We work in the $\log(E)-\log(J)$ space of the energy spectrum. The positions of knots of $x=\log(E)$ used here are 

\subsection{MCMC}

We use the Markov Chain Monte Carlo (MCMC) algorithm to find the best fiitting results, which works in the Bayesian framework. The posterior
probability of model parameters $\boldsymbol{\theta}$ is given by
\begin{equation}
p(\boldsymbol{\theta}|{\rm data}) \propto {\mathcal L}(\boldsymbol{\theta})
p(\boldsymbol{\theta}),
\end{equation}
where $p(\boldsymbol{\theta})$ is the prior probability of $\boldsymbol{\theta}$, and ${\mathcal L}(\boldsymbol{\theta})$ is the likelihood function of parameters $\boldsymbol{\theta}$ given the observational data. Here,  ${\mathcal L}(\boldsymbol{\theta}) = exp (-\chi^2/2) $, with $\chi^2$ is the defined as 
\begin{eqnarray}
\chi^2=\sum_{i=1}^{m}\frac{{\left[J(E_i;\phi_l,\phi_h,R_b)-
J_i(E_i)\right]}^2}{{\sigma_i}^2},
\end{eqnarray}
where $J(E_i;\phi_l,\phi_h,R_bi)$ is the expected flux, $J_i(E_i)$ and
$\sigma_i$ are the measured flux and error for the $i$th data bin with
geometric mean energy $E_i$.

The MCMC driver is adapted from {\tt CosmoMC} \citep{2002PhRvD..66j3511L,Liu_2012}.
We adopt the Metropolis-Hastings algorithm. The basic procedure of this
algorithm is as follows. We start with a random initial point in the 
parameter space, and jump to a new one following the covariance of these
parameters. The accept probability of this new point is defined as
$\min\left[p(\boldsymbol{\theta}_{\rm new}|{\rm data})/p(\boldsymbol{\theta}_
{\rm old}|{\rm data}),1\right]$. If the new point is accepted, then repeat
this procedure from this new one. Otherwise go back to the old point.
For more details about the MCMC one can refer to \citep{MCMC}.

\section{results and  discussion }

\begin{figure*}[!ht]
    \centering
    \includegraphics[scale=0.8]{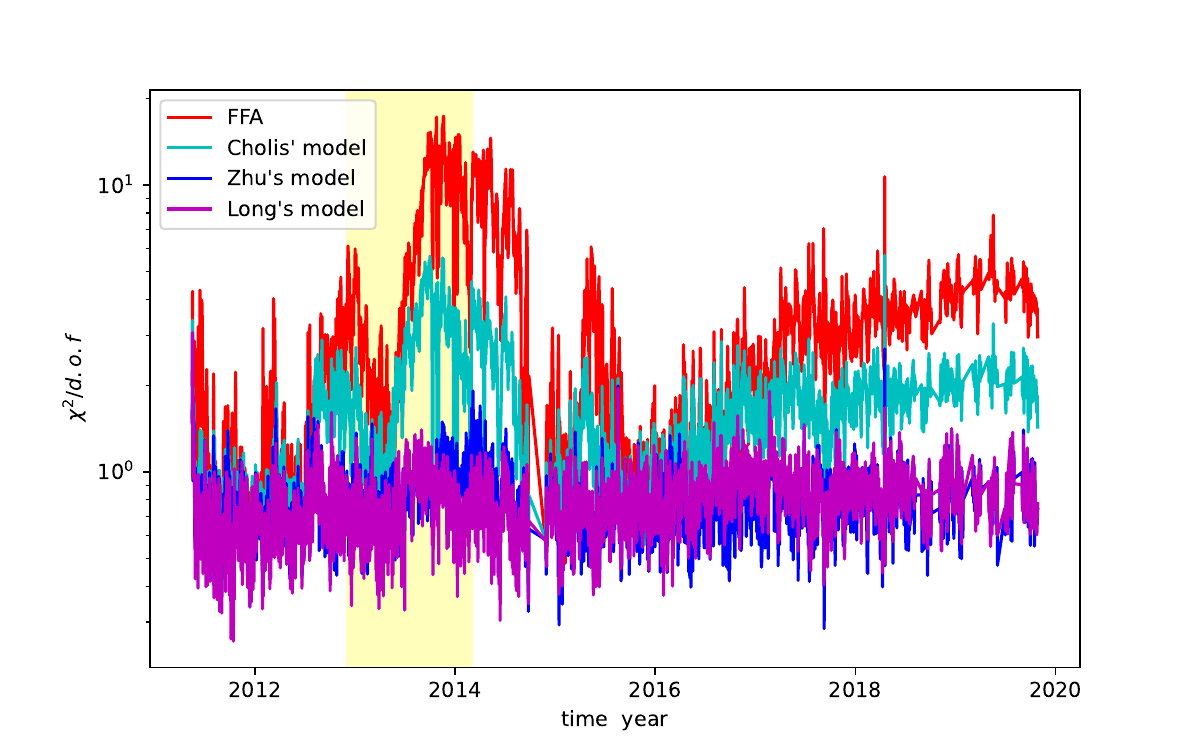}
    \caption{ The $\chi^2/d.o.f$ for the Cholis' model (cyan line), Zhu's model (red line) and Long's model (magenta line) compared to the FFA (red line). 
    The $d.o.f$ for the three modified FFA models are the same, while the $d.o.f$ for the FFA are two greater than theirs.
    The yellow band stands for the   heliospheric magnetic field reversal period within which the polarity is uncertain\citep{2015ApJ...798..114S}.}
    \label{fig:chi2}
\end{figure*}

Based on the MCMC method, we simultaneously fit the daily flux spectra of protons, Helium, and the proton-to-Helium ratio using three modified models compared to the FFA. The $\chi^2/d.o.f$ for them are shown in Fig. \ref{fig:chi2}. Consistent with previous studies, the FFA model failed to explain the time-dependent cosmic ray spectra. The FFA shows $\chi^2/d.o.f$  with values from 0.366 to 17.396 and mean value being 3.293. That's why we need a modified FFA  models to describe the solar modulation effect. 
%As demonstrated by \citep{PhysRevD.109.083009}, the  Cholis' model fits the time-dependent proton and Helium fluxes poorly, especially during the solar reversal phase and after 2016. 
As shown in \citep{PhysRevD.109.083009}, the Cholis' model fits the time-dependent proton and Helium fluxes better than the FFA, but it still has significant discrepancies during the solar polarity reversal phase and after 2016.
The $\chi^2/d.o.f$  values for the Cholis' model range from 0.367 to 5.690, with an average value of 1.629. The Zhu's model and Long's model exhibit similar fitting results, with the Long's model being slightly better. The mean $\chi^2/d.o.f$  values for Zhu's model  values for Zhu's model are 0.805, ranging from 0.284 to 3.004. There are 18 days with a $\chi^2/d.o.f$  value greater than 1.5 in the Zhu’s model, which accounts for 6.4\textperthousand \ of the data, and most of them arise around 2014 when the heliospheric magnetic field  polarity is uncertain.  
%It's interesting \textbf{to note that} the mean value of  $\chi^2/d.o.f$  is smaller than the result of \cite{zhu2024}, which is due to the low statistics, resulting in a lack of fine spectral structure in the daily data.    
The mean $\chi^2/d.o.f$  values for Long's model  are 0.793, ranging from 0.257 to 3.068, and there are 11 days with a $\chi^2/d.o.f$ value greater than 1.5 in the Long's model. Hereafter, we will present the fitting results of Zhu and Long's models.
We checked that the days with $\chi^2/d.o.f >  1.5 $  for both Zhu's and Long's models are not correlated with Forbush decrease days listed in  \cite{Wang_2023}  and in the IZMIRAN database \footnote{\url{http://spaceweather.izmiran.ru/eng/dbs.html}} .

\begin{figure*}[!ht]
    \centering
    \includegraphics[scale=0.85]{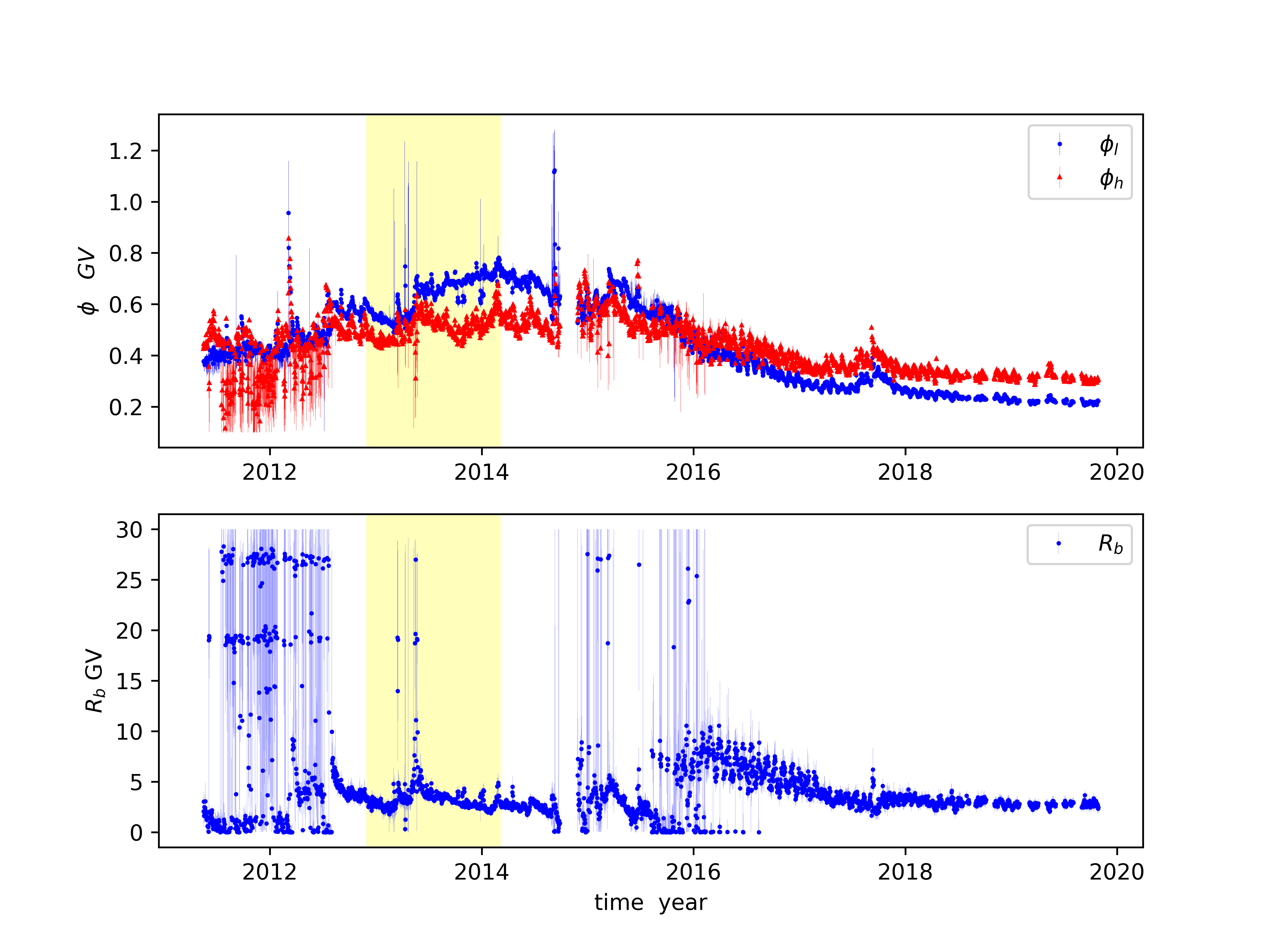}
    \caption{(Top) Time series of $\phi_l$ (blue marker)  and $\phi_h$ (red marker) via fitting to the AMS-02 p, He and p/He from 2011 to 2019. (Bottom) Same to the top but for $R_b$. 
    The yellow band stands for the heliospheric magnetic field reversal period within which the polarity is uncertain\citep{2015ApJ...798..114S}.}
    \label{fig:phi}
\end{figure*}

In the Zhu's model, the $\phi_l$, $\phi_h$ and $R_b$ are time-varying parameters to be fitted and MCMC is performed to fit the daily data of proton, Helium and p/He ratio together.  In Fig. \ref{fig:phi}, we show the fitting result, and all the specific values are provided  in Table. \ref{tab:zhu}.
%, and all the specific values are provided  in Appendix table \ref{tab:result}.
%We think the larger  $\chi^2/d.o.f$ values are induced by Forbush decreases (FDs) \citep{PhysRev.51.1108.3}. Forbush decreases (FDs)  are sudden reductions of GCRs, which happen within hours or days and are followed by a recovery period that lasts from one day to several weeks.  The causes of FDs are considered to be Interplanetary Coronal Mass Ejections (ICMEs) and Corotating Interaction Regions (CIRs). There are 146 FDs events are identified which have significant decreases in the daily flux by more than $2\sigma$  \citep{Wang_2023}. There are more FDs events are not identified as shown in the IZMIRAN database of the catalog of the Forbush-effects and interplanetary disturbances \footnote{\url{http://spaceweather.izmiran.ru/eng/dbs.html}}. The FDs will change the modulated spectra resulting in differences from the modified FFA   results,  as we not consider the FDs effect here. 
%The FDs mainly affect two aspects: increasing the  $\chi^2/d.o.f$ values or increasing the solar modulation potential. However the impact will become very small for the monthly data as shown in \cite{zhu2024}, where the $\chi^2/d.o.f$  variation is more stable.
%There are some very sharp peaks in the $\phi_l$ and $\phi_h$. The sharp peaks are also mainly due to the impact of FDs, because the FDs will decrease the fluxes as the solar modulation \citep{PhysRev.51.1108.3,2017ApJ...839...53L}. As we do not consider the FDs effect here, it will be absorbed in the solar modulation potential, causing the sharp peaks. 
The amplitude  of $\phi_h$ series  is samller than  of $\phi_l$, and it means the low rigidity particles are more sensitive to the solar activities. 
The Zhu's model is overfitting in some cases, as indicated by the large uncertainties in the fitted parameters, especially in 2012 and 2016. In these cases, we have that $\phi_l$ and $\phi_h$ are consistent with each other and/or that $R_b$ is consistent with zero, i.e., no break is effectively present. This means that when the Zhu's model overfits, it essentially reverts to the FFA behavior, as indicated also by the fact that the $\chi^2/d.o.f $ for these days is very similar between the Zhu's model and the FFA fits.

\begin{table}
\centering
\caption{The  results derived with AMS-02 daily data using Zhu's model. This table is available in its entirety in machine-readable form.} 
\begin{tabular}{ccccc}

 \toprule[1.5pt]
date & {$\phi_l$ (GV)}   & {$\phi_h$ (GV)}   & {$R_b$ (GV)}   & {$\chi^2/d.o.f$}\\
\midrule[1pt]
\midrule[1pt]

2011/5/20  &  3.704e-01$_{-3.236e-02}^{+1.396e-02}$  &  4.254e-01$_{-9.830e-03}^{+1.192e-02}$  &  2.747e+00$_{-1.329e+00}^{+1.149e+00}$  &  1.482e+00\\ 
2011/5/21  &  3.757e-01$_{-1.325e-02}^{+8.338e-03}$  &  4.489e-01$_{-1.304e-02}^{+1.499e-02}$  &  4.032e+00$_{-9.951e-01}^{+9.356e-01}$  &  3.004e+00\\ 
2011/5/22  &  3.683e-01$_{-5.727e-02}^{+1.865e-02}$  &  4.356e-01$_{-8.933e-03}^{+9.391e-03}$  &  2.242e+00$_{-1.361e+00}^{+1.056e+00}$  &  9.322e-01\\ 
2011/5/23  &  3.751e-01$_{-1.824e-02}^{+1.061e-02}$  &  4.279e-01$_{-1.011e-02}^{+1.287e-02}$  &  3.290e+00$_{-1.127e+00}^{+1.259e+00}$  &  1.835e+00\\ 
2011/5/24  &  3.563e-01$_{-3.185e-02}^{+1.608e-02}$  &  4.284e-01$_{-8.916e-03}^{+1.036e-02}$  &  2.513e+00$_{-9.765e-01}^{+9.054e-01}$  &  1.158e+00\\ 
2011/5/25  &  3.786e-01$_{-2.150e-02}^{+1.083e-02}$  &  4.314e-01$_{-1.143e-02}^{+1.450e-02}$  &  3.492e+00$_{-1.385e+00}^{+1.504e+00}$  &  1.430e+00\\ 

\bottomrule[1.5pt]
\end{tabular}
\label{tab:zhu}
\end{table}

\begin{figure*}[!ht]
    \centering
    \includegraphics[scale=0.4]{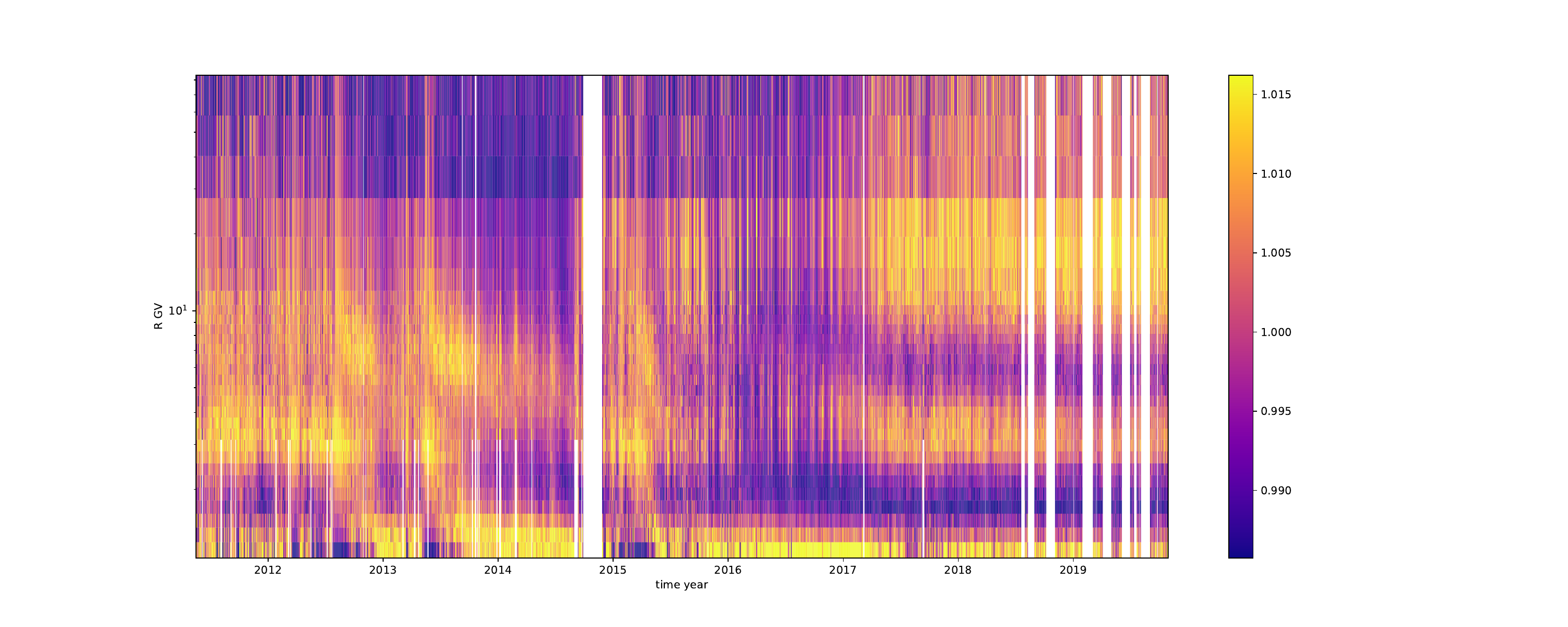}
    \caption{The ratio of Zhu's model prediction to data (model/data) for p fluxes of AMS-02  from 2011 to 2019.}
    \label{fig:p}
\end{figure*}

\begin{figure*}[!ht]
    \centering
    \includegraphics[scale=0.4]{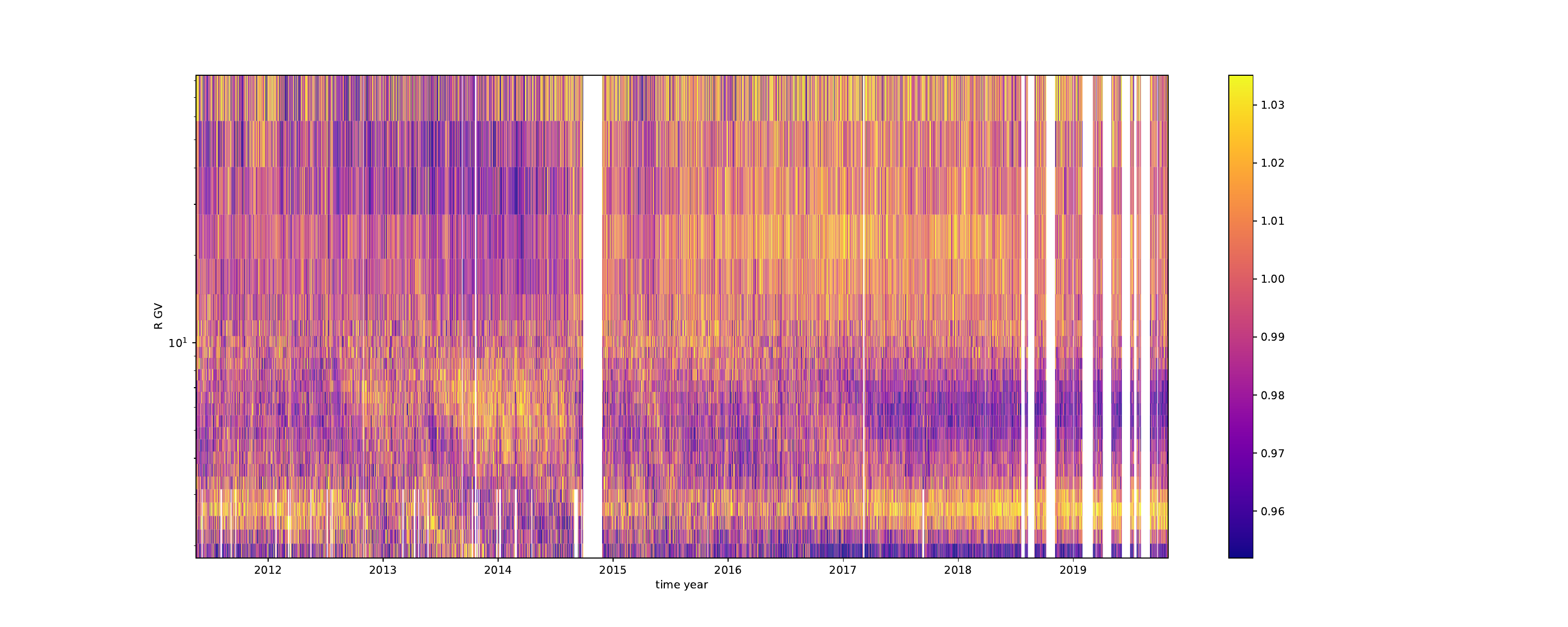}
    \caption{Same as \ref{fig:p} but for He.}
    \label{fig:He}
\end{figure*}

\begin{figure*}[!ht]
    \centering
    \includegraphics[scale=0.4]{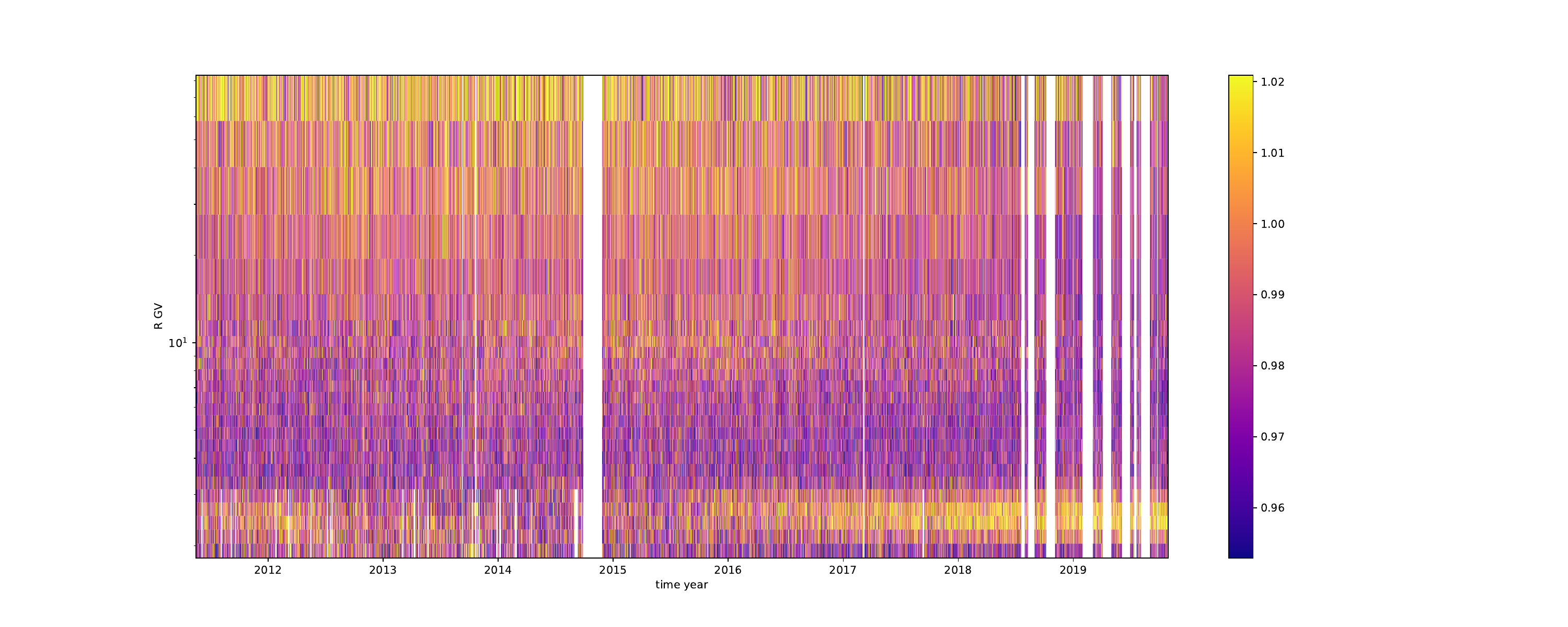}
    \caption{Same as \ref{fig:p} but for ratio of He/p.}
    \label{fig:pHe}
\end{figure*}

We show the model predictions compared to the AMS-02 daily data in  Fig. \ref{fig:p}, \ref{fig:He} and \ref{fig:pHe} from 2011 to 2019.  The figures show that the fitting agrees with the data within  $\pm\  \%5$. The fitting of proton is better, as most of the fits agree with the data within  $\pm\  \%2$. After 2017 the model predicts higher fluxes around 10 GV for both p and He.  For He, it predicts higher fluxes around 3 GV at all times, and lower fluxes at $\sim $ 2 GV after 2017. For p, there are long periods of higher predicted  fluxes at 1 GV.  These small discrepancies in the p and He fluxes are reflected in the He/p ratio, with slight systematic overpredictions at 2 GV and underpredictions at 3 GV.

% And this caused lower predicted  at around 3 GV and higher  predicted at about 2 GV in the He/p ratio. 
%The figures show our model predict less fluxes in the high energy and more fluxes in the low energy for both proton and Helium before 2017, especially during the   heliospheric magnetic field reversal period. After 2017, our model predict more fluxes in the low energy for Helium.

\begin{figure*}[!ht]
    \centering
    \includegraphics[scale=0.85]{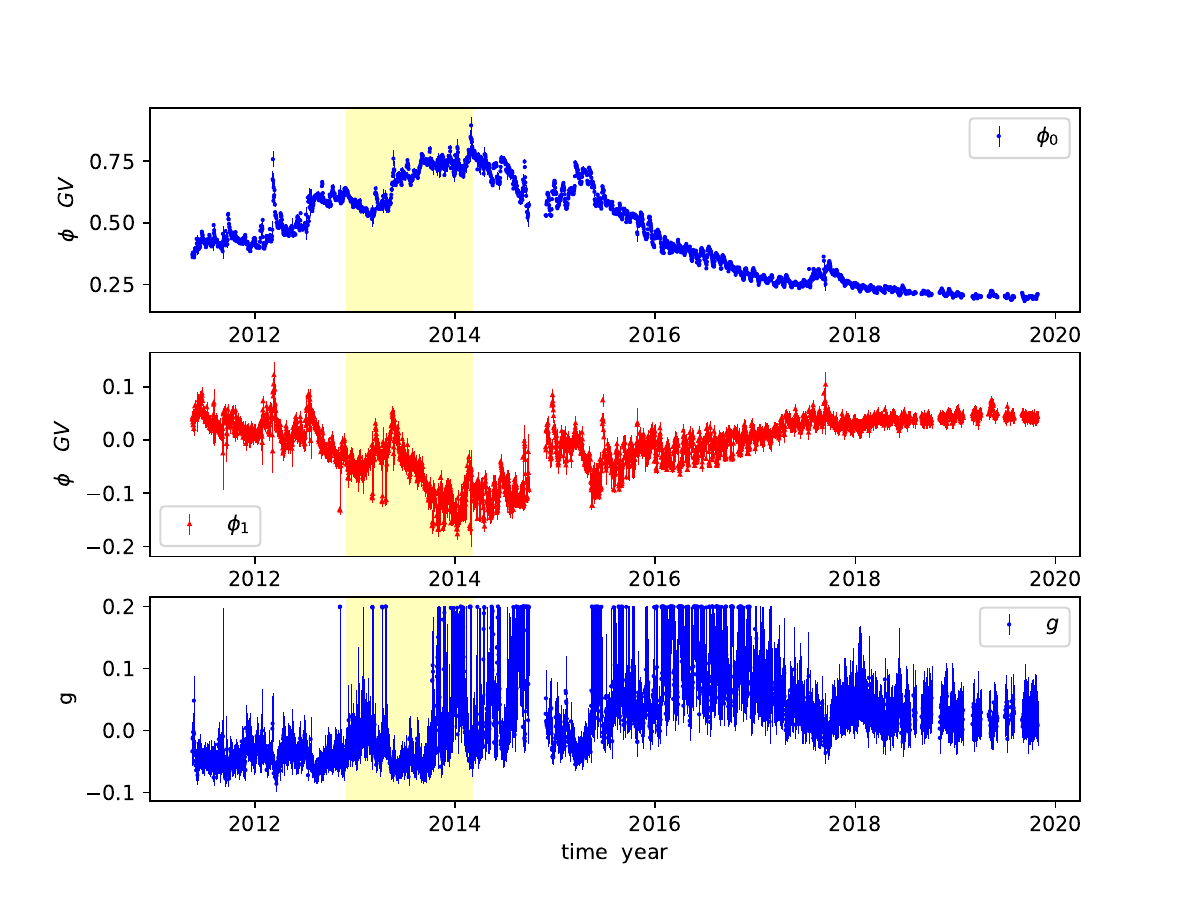}
    \caption{(Top) Time series of $\phi_0$ in Long's model via fitting to the AMS-02 p, He and p/He from 2011 to 2019. (Middle) Same to the top but for $\phi_1$. (Bottom) Same to the top but for $g$. 
    The yellow band stands for the heliospheric magnetic field reversal period within which the polarity is uncertain\citep{2015ApJ...798..114S}.}
    \label{fig:phil}
\end{figure*}

The $\phi_0$, $\phi_1$ and $g$ are the free parameters to be fitted in Long's model and the fitting results of them are shown in Fig. \ref{fig:phil}, and all the specific values are provided  in Table. \ref{tab:long}. We set the parameter $g$ range from -0.2 to 0.2 in the MCMC. The results show that the best-fit $g$ values are on the upper edge, especially in the years 2014, 2016, and 2017.
 If we expand the range of $g$, the $\chi^2/d.o.f$ will be slightly  smaller with mean value changes from 0.793 to 0.755, and $g$ will become larger. The uncertainty of these three parameters will also increase. These behaviors may be caused by the overfitting of Long's model here.

\begin{table}
\centering
\caption{The  results derived with AMS-02 daily data using Long's model. This table is available in its entirety in machine-readable form.} 
\begin{tabular}{ccccc}

 \toprule[1.5pt]
date & {$\phi_0$ (GV)}   & {$\phi_1$ (GV)}   & {$g$ }  & {$\chi^2/d.o.f$}\\
\midrule[1pt]
\midrule[1pt]
2011/5/20  &  3.715e-01$_{-7.927e-03}^{+8.047e-03}$  &  3.831e-02$_{-1.281e-02}^{+1.254e-02}$  &  -3.367e-02$_{-2.322e-02}^{+2.733e-02}$  &  1.504e+00\\ 
2011/5/21  &  3.606e-01$_{-7.352e-03}^{+7.547e-03}$  &  4.228e-02$_{-1.231e-02}^{+1.138e-02}$  &  -1.316e-02$_{-2.227e-02}^{+2.691e-02}$  &  3.068e+00\\ 
2011/5/22  &  3.795e-01$_{-7.480e-03}^{+7.410e-03}$  &  3.325e-02$_{-1.246e-02}^{+1.151e-02}$  &  -1.118e-02$_{-2.213e-02}^{+2.690e-02}$  &  9.498e-01\\ 
2011/5/23  &  3.701e-01$_{-7.146e-03}^{+7.435e-03}$  &  2.802e-02$_{-1.284e-02}^{+1.139e-02}$  &  -2.750e-03$_{-2.449e-02}^{+3.048e-02}$  &  1.855e+00\\ 
2011/5/24  &  3.619e-01$_{-7.793e-03}^{+7.739e-03}$  &  4.567e-02$_{-1.156e-02}^{+1.197e-02}$  &  -3.383e-02$_{-2.164e-02}^{+2.343e-02}$  &  1.184e+00\\ 
2011/5/25  &  3.722e-01$_{-7.445e-03}^{+7.125e-03}$  &  2.768e-02$_{-1.217e-02}^{+1.213e-02}$  &  -2.362e-03$_{-2.459e-02}^{+2.892e-02}$  &  1.439e+00\\ 
\bottomrule[1.5pt]
\end{tabular}
\label{tab:long}
\end{table}

\begin{figure*}[!ht]
    \centering
    \includegraphics[scale=0.4]{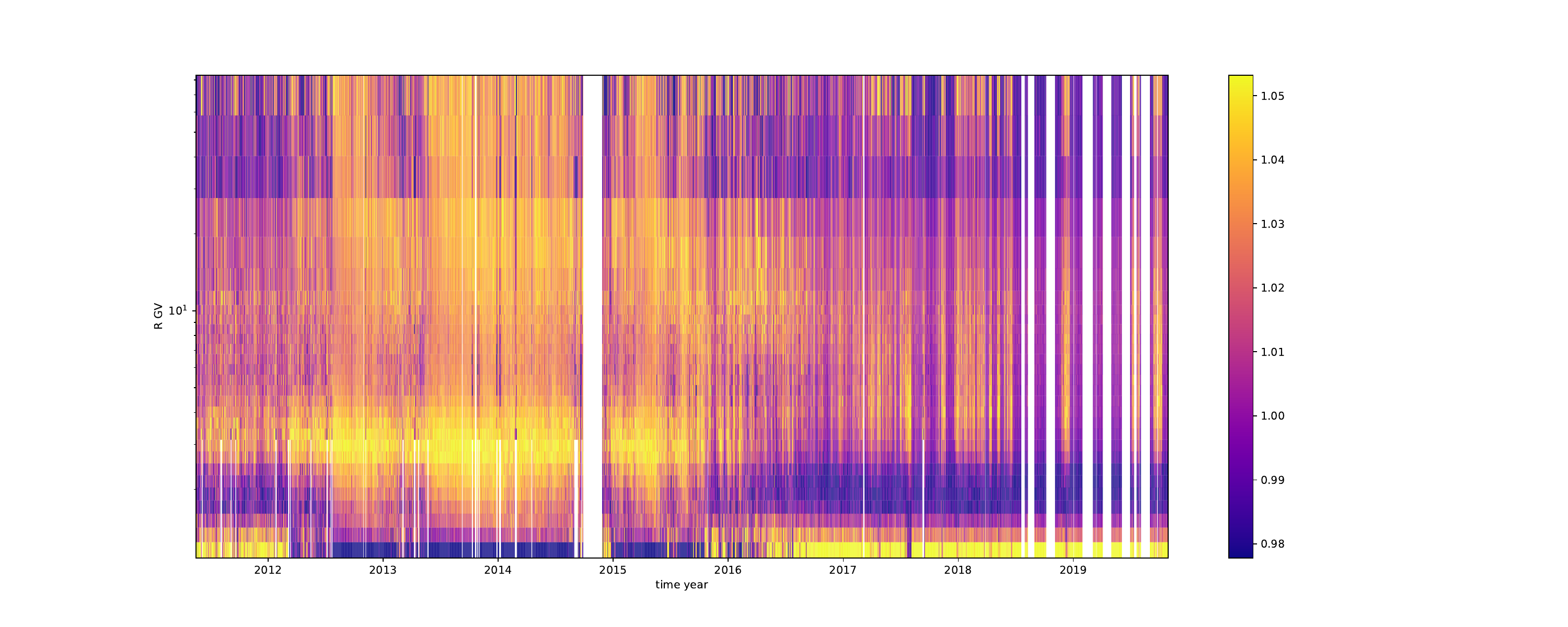}
    \caption{The ratio of Long's model prediction to data (model/data) for p fluxes of AMS-02 from  may 2011 to may 2019 }
    \label{fig:pl}
\end{figure*}

\begin{figure*}[!ht]
    \centering
    \includegraphics[scale=0.4]{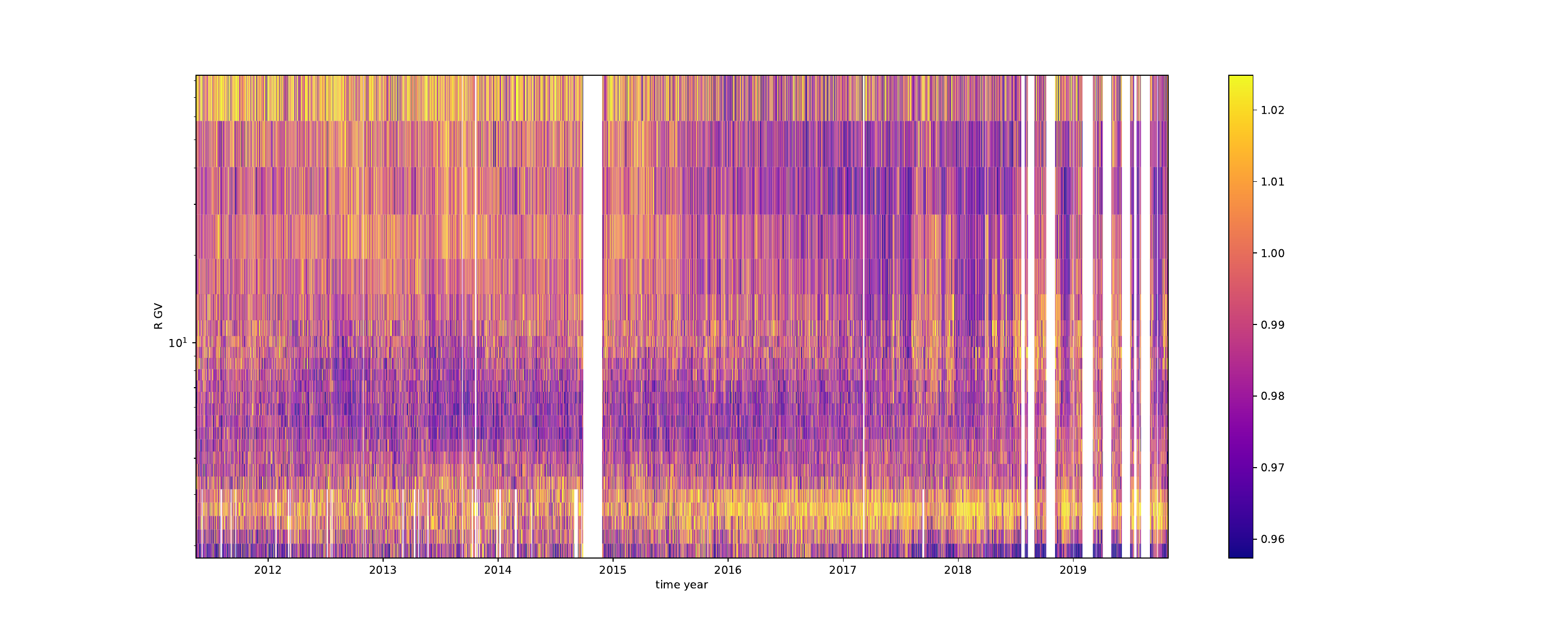}
    \caption{Same as \ref{fig:pl} but for He.}
    \label{fig:Hel}
\end{figure*}

\begin{figure*}[!ht]
    \centering
    \includegraphics[scale=0.4]{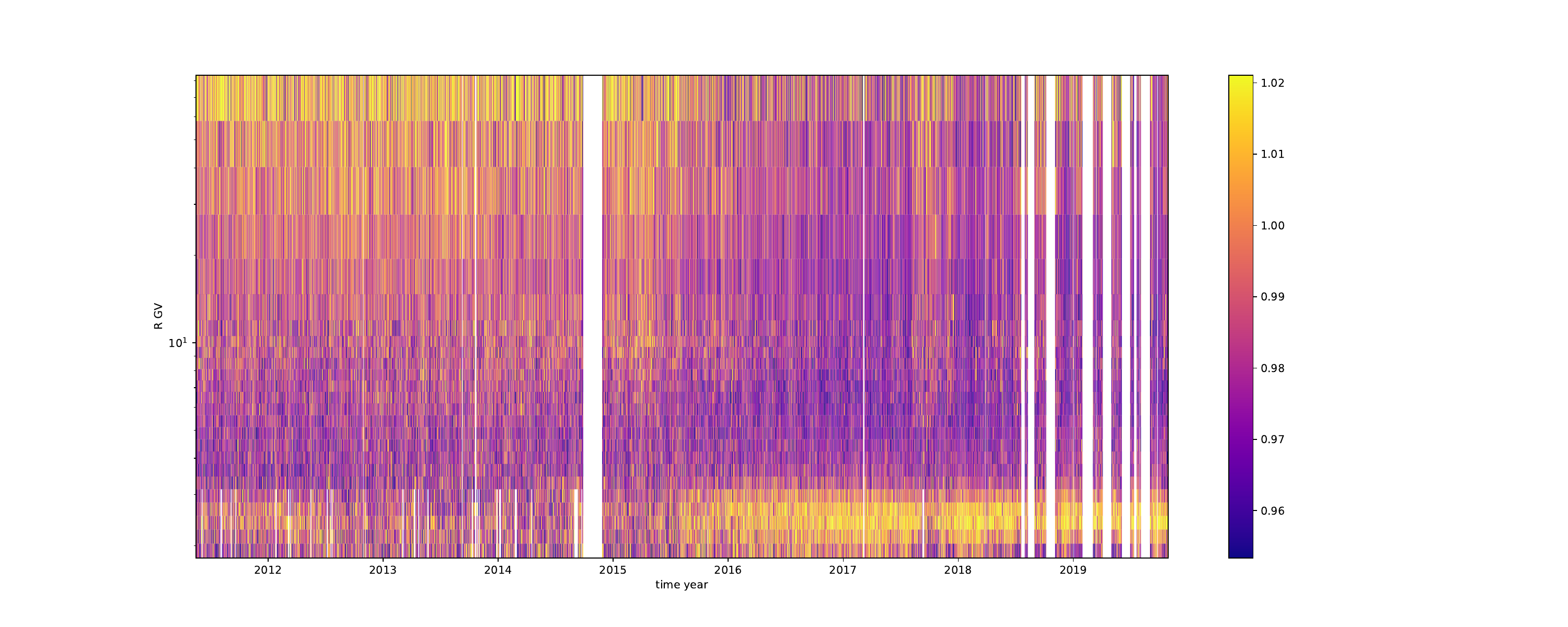}
    \caption{Same as \ref{fig:pl} but for ratio of He/p.}
    \label{fig:pHel}
\end{figure*}

The Long's model predictions compared to the AMS-02 p, He and p/He ratio data from 2011 to 2019 are shown in Fig. \ref{fig:pl}, \ref{fig:Hel} and \ref{fig:pHel}. Similar to the Zhu's model, most of the predictions agree with the data within $\pm$5\%. The behavior of model/data is also similar to Zhu's. Except for the proton fluxes, Long's fitting results are slightly worse than Zhu's. And there are long periods of smaller predicted fluxes before 2017 and high predicted fluxes after that at 1 GV for p.

\begin{figure*}[!ht]
    \centering
    \includegraphics[scale=0.65]{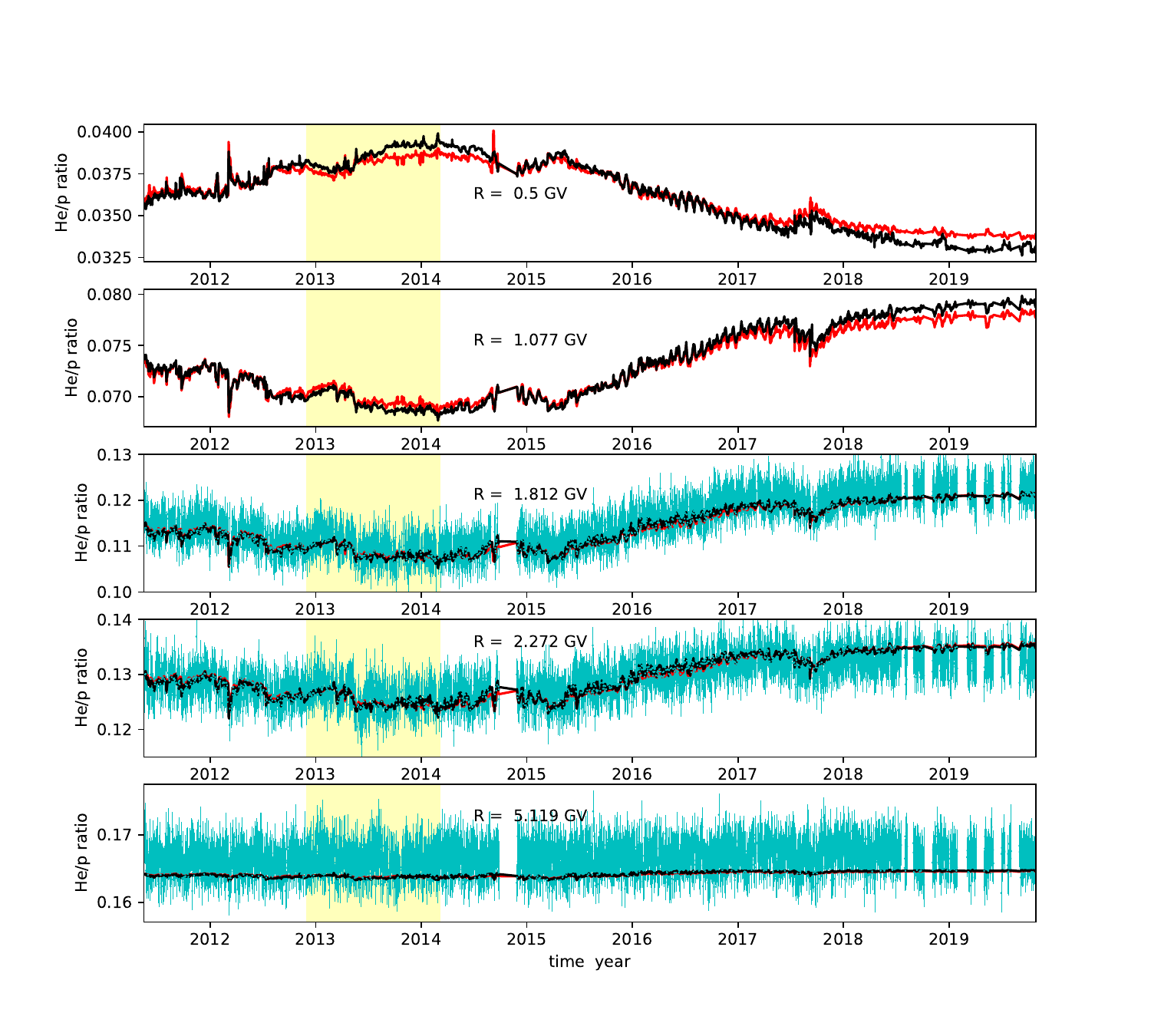}
    \caption{The best-fit time profiles of the Helium-to-proton ratio from Zhu's model (red)  and Long's model (black) evaluated at rigidities R = 0.5 GV, 1.077 GV, 1.812 GV, 2.272 GV, and 5.119 GV (from top to bottom) are compared with the AMS-02 data (cyan).}
    \label{fig:ratio}
\end{figure*}

In Fig. \ref{fig:ratio}, we  present the time profile of the Helium-to-proton ratio (He/p) for several rigidity values from Zhu and Long's model. Using the same modulation parameters for both p and He, the ratio of  He/p  can be fitted very well. The two models both successfully reproduced the time evolution of He/p from 2011 to 2019. 
We show the He/p ratio at R = 0.5 GV and 1.077 GV where there is no ASM-02 data. 
At  R = 0.5 GV,  Long's model predicts a higher He/p ratio than Zhu's model in 2014, and a lower ratio after 2017. At R = 1.077 GV, the situation is the opposite.
At higher rigidities, Long's model prediction is very similar to Zhu's model prediction.
 At the rigidity R = 0.5 GV, the He/p increase from 2011 to 2014 with increasing of solar activity, and then decrease to low values in 2014–2019.  At higher rigidities, the behavior is completely opposite. At even higher rigidities, the He/p ratio will not change with time.
The long-term behavior of  the He/p ratio at the low rigidities is caused by two reasons: the different Z/A value and LIS for p and He as discussed in our previous work \citep{zhu2024} as they share the same solar modulation parameters.  For more information about this one can refer to \cite{zhu2024}.

%Although, at very low rigidities such as R = 1.812 GV, the model's prediction is slightly lower than the data, especially after 2017. This may be cause by the  hysteresis between fluxes of p and He. However, the hysteresis effects here are very slight, as the model's predicted ratio agrees with the observational data within error bands.

\section{conclusion }

The precise measurements of daily cosmic ray proton and Helium
spectra by AMS-02 between 2011 to 2019 provide an important
chance to improve our understanding for the solar modulation. In this paper, we fit the daily  p, He and p/He data from AMS-02 with three modified FFA models and FFA model.  
%With  the non-parametric LIS of proton and Helium from \cite{zhu2024}, we found with a sigmoid function to replace the solar modulation potential in  the classic FFA, we can fit the data very well.
We find that, using the non-parametric LIS from \cite{zhu2024} for proton and Helium, we can fit the data very well with two modified FFA models, Zhu's model and Long's model.
In the Zhu's model, a sigmoid function is used to replace the constant $\phi$ in the FFA.
The sigmoid function has three parameters, $\phi_l$ for low rigidity, $\phi_h$ for high rigidity and $R_b$ for the break rigidity. 
%If the best fit  $R_b$ is too low,  rigidity-dependent of  $\phi(R)$ here will close to the result of \citep{2017JGRA..12210964G} in the low energy. The time evolution of $\phi_h$ is similar to the result from NM, as they are  both associated with the high energy particles. Due to the different assumption of LIS, the  values of  $\phi_h$  are 0.17 GV smaller than the result from NM. 
The evolution of $\phi_l$ has a larger amplitude than $\phi_h$, which means the low energy particles are more sensitive to the solar activities. In the Zhu's model, if $\phi_l$ is close to $\phi_h$, $R_b$ is too high or too low, the modified FFA model will be overfitting, and transform into FFA.
There are also three parameters in Long's model: $\phi_0$ and $\phi_1$ to describe the linear relation between $\phi(R)$ and $\ln (R)$, while $g$ controls how much the drift processes affect the modulated spectrum. The model seems to be overfitting in 2014, 2016 and 2017, but it does not transform into FFA.

 The Zhu's model and Long's model show very similar fitting results with mean $\chi^2/d.o.f$ values of approximately 0.8, which are much better than the other two models. The two models prediction are very similar to each other,  especially for the He/p ratio. The  predictions both agree with the data within $\pm$ 5\%, using the same parameters fitted with both protons and Helium simultaneously. Therefore, the data are consistent with the assumption that p and He undergo the same propagation processes in the heliosphere.

%In this work, we can fit the proton and Helium with the same parameters, which means the propagation process of proton and Helium is  the same. The short-term fluctuations (e.g. FDs) are significant behavior in the daily data, which are associated with the \textbf{space weather} events such as ICMEs. Our model is able to give a well fitting for the mean  $\chi^2/d.o.f = 0.778$,  which  means the short-term behavior may be absorbed by the solar modulation potential. 
%Our model cannot reproduce the slight hysteresis between the fluxes of p and He, which may be caused by the drift effect or other Z/A-dependent effects that are not considered in this model. We may need more realistic modulation models (such as stochastic differential equations \citep{2016CoPhC.207..386K,2017ApJ...839...53L,2022PhRvL.129w1101Z}) for further research in the future.

%Positron and proton fluxes exhibit similar long-term behavior \citep{PhysRevLett.131.151002}, and there is a nearly linear correlation between the fluxes of $e^+$ and p throughout the entire time period. With proper assumption of LIS, $e^+$ and p may also share the same solar modulation parameters with our model as shown in \cite{Zhu:2020koq}, and we will study this in the future.

%The well fitting results of Zhu's and Long's model in this work verify the validity and correctness of the MFFA  with the daily fluxes. 
The results in this study verify that the modified FFA model is a valid parametrization of the GCR spectrum also at daily time scales.
Usually, the FFA combined with GALPROP \footnote{\url{http://galprop.stanford.edu/}} \citep{Strong_1998,Moskalenko_1998} is used to constrain the origin and propagation of GCRs in the galaxy. In order to get better fitting results, different solar modulation parameters are adopted for different CR species \citep{Yuan_2018}. 
Actually, We need a sophisticated 3D time-dependent self-consistent model \citep{1999ApJ...513..409Z} to fully solve the problem related to cosmic ray transport in the heliosphere \citep{Potgieter2013}. However it is too complex and computationally intensive to  cooperate it with  combination  of GALPROP and MCMC \citep{2017PhRvD..95h3007Y,2017ApJ...840..115B}. 
If we consider all the GCRs which have the same charge sign share the rigidity-dependent solar modulation potential in the study of propagation of GCRs in the galaxy, it will be useful to obtain the propagation parameters, as well as search for new physics such as dark matter signals in the antiproton flux.
This can be tested by checking if the Zhu's and Long's models of modified FFA can reproduce daily AMS-02 positron data using the same modulation parameters as found by fitting p and He daily data. If so, the modulation parameters derived from daily AMS-02 electron data could be used to predict the time variation of anti-protons. This will be a subject of future studies.

%A full solution to the problem related to CR transport in the heliosphere is a complicated task and requires sophisticated 3D time-dependent self-consistent modelling. 

%%%%%%%%%%%%%%%%%%%%%%%%%%%%%%%%%%%%%%%%%%%%%%%%%%%%%%%%%%%%%%
\begin{acknowledgments}
Thanks for Qiang  Yuan for very helpful discussions. This work is supported by the National Natural Science Foundation of China  (No. 12203103). C.R.Z is also support by the Doctoral research start-up funding of Anhui Normal University. We acknowledge the use of the data from the \href{https://tools.ssdc.asi.it/CosmicRays/}{Cosmic-Ray Database (https://tools.ssdc.asi.it/CosmicRays/)} \citep{DiFelice:2017Hm}. %\footnote{\url{https://tools.ssdc.asi.it/CosmicRays/}}.
%and {Cosmic-Ray Data Base (CRDB)}\citep{Maurin:2023alp} \footnote{\url{https://lpsc.in2p3.fr/crdb/}}. 
\end{acknowledgments}
%%%%%%%%%%%%%%%%%%%%%%%%%%%%%%%%%%%%%%%%%%%%%%%%%%%%%%%%%%%%%%
%\setcounter{figure}{0}
%\renewcommand\thefigure{A\arabic{figure}}
%%%%%%%%%%%%%%%%%%%%%%%%%%%%%%%%%%%%%%%%%%%%%%%%%%%%%%%%%%%%%%
\clearpage

%%%%%%%%%%%%%%%%%%%%%%%%%%%%%%%%%%%%%%%%%%%%%%%%%%%%%%%%%%%%%%
%\setcounter{figure}{0}
%\renewcommand\thefigure{A\arabic{figure}}

\bibliographystyle{aasjournal}
\bibliography{sample631}{}

%\bibliographystyle{apsrev}
%\bibliography{refs}

\end{document}